\PassOptionsToPackage{table, dvipsnames}{xcolor}
\documentclass[sigconf]{acmart}

\makeatother

\usepackage{xcolor,colortbl}
\usepackage{stackengine}
\usepackage[flushleft]{threeparttable}

\usepackage{bm}
\usepackage{multirow}
\usepackage{graphicx}
\usepackage{algorithm}
\usepackage[noend]{algpseudocode}
\usepackage[frozencache=true,cachedir=minted-cache,newfloat]{minted}

\usepackage{hyperref}
\usepackage{adjustbox}
\usepackage{multirow}
\usepackage{xspace}
\usepackage{soul}
\sethlcolor{green}

\usepackage{cleveref}
\usepackage{pifont}

\usepackage{enumitem}
\usepackage{pgfplots}
\usepackage{makecell}
\usepackage{diagbox}
\usepackage{svg}

\definecolor{lightgreen}{RGB}{200,255,200}

\usepgfplotslibrary{colorbrewer}
\usetikzlibrary{calc,shadings,patterns}

\pgfplotsset{compat=1.18}

\renewcommand\footnotetextcopyrightpermission[1]{}
\copyrightyear{2024}
\acmYear{2024}
\setcopyright{rightsretained}
\acmConference[WWW '24 Companion]{Companion Proceedings of the ACM Web
Conference 2024}{May 13--17, 2024}{Singapore, Singapore}
\acmBooktitle{Companion Proceedings of the ACM Web Conference 2024 (WWW
'24 Companion), May 13--17, 2024, Singapore,
Singapore}\acmDOI{10.1145/3589335.3651440}
\acmISBN{979-8-4007-0172-6/24/05}

\makeatletter
\gdef\@copyrightpermission{
 \begin{minipage}{0.3\columnwidth}
  \href{https://creativecommons.org/licenses/by-nc-sa/4.0/}{\includegraphics[width=0.90\textwidth]{figure/4ACM-CC-by-nc-sa-88x31.eps}}
 \end{minipage}\hfill
 \begin{minipage}{0.7\columnwidth}
  \href{https://creativecommons.org/licenses/by-nc-sa/4.0/}{This work is licensed under a Creative Commons Attribution-NonCommercial-ShareAlike International 4.0 License.}
 \end{minipage}
 \vspace{5pt}
}
\makeatother

\begin{document}
\title[\framework: Towards a More Up-to-Date Unified Framework for the Extraction of Multimodal Feat. in RecSys]{\framework: Towards a More Up-to-Date Unified Framework for
the Extraction of Multimodal Features in Recommendation}

\author{Matteo Attimonelli}
\affiliation{\institution{Politecnico di Bari, Italy}
  \city{}
  \country{}}
\email{matteo.attimonelli@poliba.it}

\author{Danilo Danese}
\affiliation{\institution{Politecnico di Bari, Italy}
  \city{}
  \country{}}
\email{danilo.danese@poliba.it}

\author{Daniele Malitesta}
\authornote{Work done while at Politecnico di Bari.}
\affiliation{\institution{Université Paris-Saclay, CentraleSupélec, Inria, France}
  \city{}
  \country{}}
\email{daniele.malitesta@centralesupelec.fr}

\author{Claudio Pomo}
\affiliation{\institution{Politecnico di Bari, Italy}
  \city{}
  \country{}}
\email{claudio.pomo@poliba.it}

\author{Giuseppe Gassi}
\affiliation{\institution{Politecnico di Bari, Italy}
  \city{}
  \country{}}
\email{g.gassi@studenti.poliba.it}

\author{Tommaso {Di Noia}}
\affiliation{\institution{Politecnico di Bari, Italy}
  \city{}
  \country{}}
\email{tommaso.dinoia@poliba.it}

\renewcommand{\shortauthors}{Matteo Attimonelli et al.}

\def\framework{\textsc{Ducho 2.0}\xspace}
\def\ducho{\textsc{Ducho 2.0}\xspace}
\def\duchoOld{\textsc{Ducho}\xspace}


\newcommand{\matteo}[1]{\textcolor{red}{{\bf [Matteo: }{\em #1}{\bf ]}}}
\newcommand{\danilo}[1]{\textcolor{teal}{{\bf [Danilo: }{\em #1}{\bf ]}}}
\newcommand{\daniele}[1]{\textcolor{blue}{{\bf [Daniele: }{\em #1}{\bf ]}}}
\newcommand{\claudio}[1]{\textcolor{brown}{{\bf [Claudio: }{\em #1}{\bf ]}}}





\begin{abstract}
In this work, we introduce \framework, the latest stable version of our framework. Differently from \duchoOld, \framework offers a more personalized user experience with the definition and import of custom extraction models fine-tuned on specific tasks and datasets. Moreover, the new version is capable of extracting and processing features through multimodal-by-design large models. Notably, all these new features are supported by optimized data loading and storing to the local memory. To showcase the capabilities of \framework, we demonstrate a complete multimodal recommendation pipeline, from the extraction/processing to the final recommendation. The idea is to provide practitioners and experienced scholars with a ready-to-use tool that, put on top of any multimodal recommendation framework, may permit them to run extensive benchmarking analyses. All materials are accessible at: \url{https://github.com/sisinflab/Ducho}.
\end{abstract}

\maketitle

\section{Introduction and Motivations}

Multimodal data sources (e.g., product images, descriptions, reviews, audio tracks) support recommendation systems in tasks such as fashion, micro-videos, and food recommendation. Indeed, the extraction of meaningful multimodal features from such data empowers the recommendation models by enriching their knowledge and understanding of users' preferences, eventually improving the quality of the proposed personalized suggestions. Nevertheless, to date, no standardized solutions for multimodal feature extraction/processing still exist in multimodal recommendation.


In our work~\cite{DBLP:conf/mm/MalitestaGPN23}, we introduced \duchoOld as a solution to unify the extraction and processing of multimodal features in recommendation systems. \duchoOld facilitates the creation of a comprehensive multimodal feature extraction pipeline, allowing users to specify data sources, backends, and deep learning models for each modality. The pipeline is easily configurable through a YAML file, simplifying the extraction and processing organized as sub-modules.

Even if the current functionalities enable to perform sufficiently extensive multimodal extraction pipelines to power the majority of existing multimodal recommender systems, we still recognize room for improvement in terms of the usability and optimization of the framework. Moreover, in the ever-evolving landscape of multimodal deep learning and recommendation, and (especially) with the recent outbreak of large models trained for several deep learning tasks, it becomes imperative to keep \duchoOld always up-to-date to implement and reflect such advances also in our framework. 


Motivated by the outlined aspects, we present \framework, the latest stable version succeeding \duchoOld. Our contributions focus on two main aspects: (i) improving the framework's \textbf{usability} and \textbf{customization} with \textbf{optimized} procedures, and (ii) enhancing multimodal feature extraction by incorporating recent advancements in \textbf{large multimodal models}.

Regarding the (i) contribution, \framework supports the adoption of custom extractor models with custom extraction layers, and the application of pre-processing operations (e.g., image normalization) whose parameters may be easily modified depending on the user's needs. Furthermore, towards efficiency, we introduce the popular PyTorch custom dataloader, that helps to optimize the overall loading/storing process. As for the (ii) contribution, we enrich the set of available backends-modalities configurations and implement the extraction/process of multimodal features through multimodal-by-design large models (such as CLIP~\cite{DBLP:conf/icml/RadfordKHRGASAM21}). In this respect, \framework is also capable of performing multimodal fusion operations.

The reminder of this paper elucidates each of the novelties introduced with \framework. Additionally, we also propose a demonstration to showcase how \framework may be seamlessly used within a complete multimodal recommendation pipeline, on top of the framework Elliot~\cite{DBLP:conf/sigir/AnelliBFMMPDN21} tailored to address multimodal recommendation~\cite{DBLP:journals/corr/abs-2309-05273}. Indeed, we hope to provide practitioners and experienced scholars with an easy-to-use tool to run extensive benchmarking analyses with multimodal recommender systems, thus opening novel possible directions in the field. We release all the useful material for \framework at: \url{https://github.com/sisinflab/Ducho}.

\begin{figure*}[!t]
\center
\includegraphics[width=\textwidth]{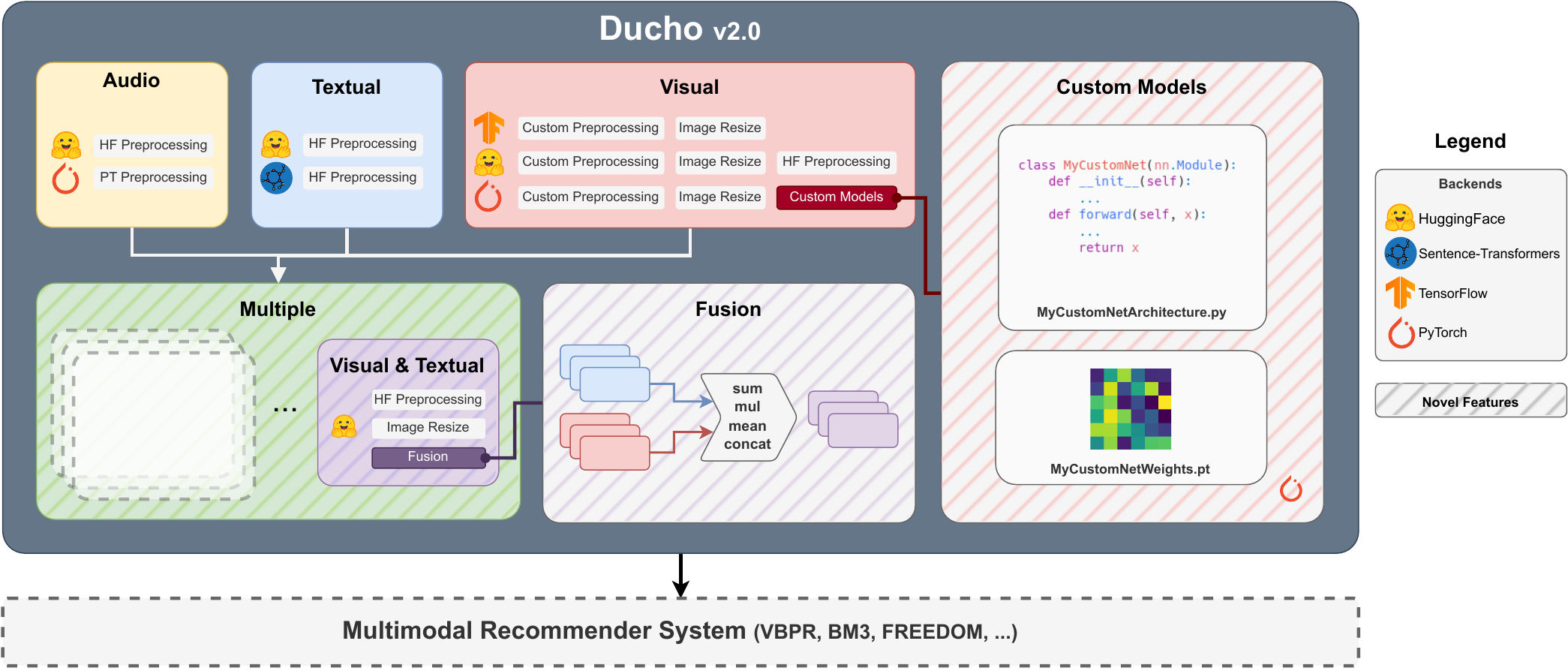}
    \caption{An overview of \framework, where newly-introduced functionalities have hatch background.}
    \label{fig:ducho_overview}
\end{figure*}

\section{Novel features}

This section delves into \framework, presenting new functionalities compared to \duchoOld~\cite{DBLP:conf/mm/MalitestaGPN23}. The introduced features are categorized into two categories: (i) customization and optimization, and (ii) backends and large multimodal models. \Cref{fig:ducho_overview} illustrates the architecture of \framework, emphasizing the newly-added features.

\subsection{Customization and optimization}

\noindent $\bullet$ \textbf{Pipeline optimization.} \framework aims at enhancing computational efficiency. One significant addition is the implementation of multiprocessing facilitated by PyTorch-based dataloaders. Indeed, this new feature effectively leads to faster data loading and storing, which may represent a crucial bottleneck in the overall performance. Moreover, \framework is now capable of leveraging the computational speedup of MPS technology that alongside CUDA (already available in the previous version) makes our framework suitable across several existing platforms.

\noindent $\bullet$ \textbf{More flexibility and personalization.} \framework introduces a suite of novel features aimed at providing users with more customization options for the multimodal extraction pipeline. Noteworthy additions include the ability to specify desired types of pre-processing operations on images (i.e., z-score or minmax normalizations) that can be further customized through specific parameters. Moreover, users may now perform extractions using custom PyTorch and Transformers models for the \textcolor{Mahogany}{\textbf{visual}} and \textcolor{RoyalBlue}{\textbf{textual}} modalities, enhancing the flexibility and adaptability of the framework. In this respect, \framework also offers multiple image processors and tokenizers (which may be designed and pre-trained by the user) adding another useful level of customization.

\subsection{Backends and large multimodal models}

\noindent $\bullet$ \textbf{Additional backends settings.} \framework introduces advancements regarding backends. Notably, the Transformers backend is now available also for the \textcolor{Mahogany}{\textbf{visual}} modality together with the \textcolor{RoyalBlue}{\textbf{textual}} one (already present in the previous framework version). This novel feature comes as fundamental to enrich the extraction models' collection available in \framework through the extensive hub of pre-trained networks available on HuggingFace. 

\noindent $\bullet$ \textbf{Multiple modalities and fusion.} Following the recent advances in large multimodal models, \framework now opens to the extraction and processing of multimodal features by directly using multimodal-by-design networks (e.g., CLIP \cite{DBLP:conf/icml/RadfordKHRGASAM21}). Specifically, the framework integrates the \textbf{\textcolor{RoyalPurple}{visual\_textual}} modality, building the necessary implementation bases for future extensions with other combined modalities. Then, users can require unified representations of the generated multimodal features by fusing their embeddings through various popular methods, such as concatenation, element-wise summation/multiplication, and averaging.

\section{The new configuration file}

To provide a more technical description of \framework, we show a toy YAML configuration file (Configuration~\ref{lst:ducho_v2_yaml}) which spans all the core functionalities introduced in \framework. Note that, to support the seamless transaction from \duchoOld to \framework, the configuration file retains the same modular structure as before.

First, in the \textbf{\textcolor{Mahogany}{visual}} extraction, users can set the \texttt{preprocessing} procedure to perform (i.e., either z-score or minmax, optionally with custom \texttt{mean} and \texttt{std} values) in the case of PyTorch backend. Under the same backend setting, the configuration may also include the loading of a custom extraction model; note that the user will need to indicate the exact path to the pre-trained weights (i.e., the \texttt{.pt} file) and include or import the definition of its architecture as a \texttt{torch.nn.Model} in the main Python script. Similarly, when leveraging the Transformers backend, it is now possible to specify a custom \texttt{model\_name}, as the framework will first search on the HuggingFace hub and (if not available) at a local path, allowing to utilize custom pre-trained models. Noteworthy, the \texttt{image\_processor} and the \texttt{output\_layers} may be also customized. 

Then, the \textbf{\textcolor{RoyalBlue}{textual}} extraction permits to load pre-trained tokenizers from HuggingFace or rely on own specified tokenizer using the \texttt{tokenizer\_name} parameter. In addition to this, we have fixed some code bugs regarding the names of the \texttt{.tsv} columns standing for the items' IDs and their description; thus, it is now possible to explicitly indicate custom column identifiers.  

Finally, in the novel scenarios involving multiple modalities (e.g., \textbf{\textcolor{RoyalPurple}{visual\_textual}}), it is worth underlining that fusion techniques are also available to combine the extracted \textbf{\textcolor{Mahogany}{visual}} and \textbf{\textcolor{RoyalBlue}{textual}} modalities into a single representation. Currently, \framework supports concatenation (i.e., \texttt{concat}), as well as element-wise operations such as summation (i.e., \texttt{sum}), multiplication (i.e., \texttt{mul}), and average (i.e., \texttt{mean}). Note that, in such cases, the framework will check for dimensionalities mismatch, as for element-wise operations the two fused vectors are supposed to have the same embedding size. Moreover, output folders for the fused multimodal features will be named accordingly (e.g., \texttt{vis\_embeddings\_text\_embeddings\_concat}). 

\setlength{\fboxsep}{1.3pt}

\begin{listing}[!t]
\begin{minted}[
    frame=lines,
    framerule=0.8pt,
    escapeinside=||,
    fontsize=\footnotesize,
  ]{text}

|\textbf{\textcolor{Black}{dataset\_path}}|: ./my/dataset/path
|\textbf{\textcolor{Black}{gpu list}}|: 0
|\textbf{\textcolor{Mahogany}{visual}}|:
 |\textbf{\textcolor{Black}{items}}|:
  |\textbf{\textcolor{Black}{input\_path}}|:  images
  |\textbf{\textcolor{Black}{output\_path}}|: visual_embeddings
  |\textbf{\textcolor{Black}{model}}|: [
    { |\textbf{model\_name}|: ResNet18, |\textbf{output\_layers}|: avgpool,
      |\textbf{reshape}|: [224, 224], |\textbf{backend}|: torch, |\colorbox{lightgreen}{\textbf{preprocessing}}|: zscore,
      |\colorbox{lightgreen}{\textbf{mean}}|: [0.485, 0.456, 0.406], |\colorbox{lightgreen}{\textbf{std}}|: [0.229, 0.224, 0.225] },
    { |\textbf{model\_name}|: |\colorbox{lightgreen}{./MyCustomNetWeights.pt}|, |\textbf{backend}|: torch,
      |\textbf{output\_layers}|: pooler_output, |\colorbox{lightgreen}{\textbf{preprocessing}}|: minmax },
    { |\textbf{model\_name}|: |\colorbox{lightgreen}{./MyCustomHFModel}|, |\textbf{backend}|: |\colorbox{lightgreen}{transformers}|,
      |\textbf{output\_layers}|: |[\colorbox{lightgreen}{MyCustomOutputLayer}, avgpool]|,
      |\colorbox{lightgreen}{\textbf{image\_processor}}|: ./MyCustomImageProcessor } ]     
|\textbf{\textcolor{RoyalBlue}{textual}}|:
 |\textbf{\textcolor{Black}{items}}|:
  |\textbf{\textcolor{Black}{input\_path}}|:  descriptions.tsv
  |\textbf{\textcolor{Black}{output\_path}}|: textual_embeddings
  |\textbf{\textcolor{Black}{item\_column}}|: |\colorbox{lightgreen}{asin}|
  |\textbf{\textcolor{Black}{text\_column}}|: |\colorbox{lightgreen}{description}|
  |\textbf{\textcolor{Black}{model}}|: [
   { |\textbf{model\_name}|: |\colorbox{lightgreen}{./MyCustomHFModel}|, |\textbf{clear\_text}|: False,
     |\textbf{output\_layers}|: |\colorbox{lightgreen}{MyCustomOutputLayer}|, |\textbf{backend}|: transformers, 
     |\colorbox{lightgreen}{\textbf{tokenizer\_name}}|: |./MyCustomTokenizer| } ] 
|\colorbox{lightgreen}{\textbf{\textcolor{RoyalPurple}{visual\_textual}}}|:
 |\textbf{\textcolor{Black}{items}}|:
  |\textbf{\textcolor{Black}{input\_path}}|:  { |\textbf{visual}|: images, |\textbf{textual}|: meta.tsv }
  |\textbf{\textcolor{Black}{output\_path}}|: { |\textbf{visual}|: vis_embeddings, |\textbf{textual}|: text_embeddings }
  |\textbf{\textcolor{Black}{item\_column}}|: asin
  |\textbf{\textcolor{Black}{text\_column}}|: description
  |\textbf{\textcolor{Black}{model}}|: [
   { |\textbf{model\_name}|: openai/clip-vit-base-patch16, |\colorbox{lightgreen}{\textbf{fusion}}|: concat, 
     |\textbf{output\_layers}|: 1, |\textbf{backend}|: transformers  } ]
\end{minted}
\caption{A toy example with the YAML configuration. New features for \framework are highlighted in \colorbox{lightgreen}{green}.}
\label{lst:ducho_v2_yaml}
\end{listing}

\section{Multimodal Benchmarks}


This section highlights \framework into a multimodal recommendation pipeline, demonstrating its usability as a feature extractor for benchmarking any multimodal recommender system. We show two new functionalities: (i) feature extraction through multimodal-by-design models, and (ii) custom models. We outline the construction of the multimodal dataset, describe the feature extraction, present the multimodal recommendation approaches, explain how to run the demonstration and discuss the results.

\subsection{Dataset preparation}

We use \textbf{Amazon Baby} from the Amazon recommendation dataset\footnote{\url{https://cseweb.ucsd.edu/~jmcauley/datasets/amazon/links.html}.}. Recorded items are enriched with metadata, with their unique identifiers, textual descriptions, and image URLs for product photos. In the preprocessing phase, we retain products with valid image URLs and non-empty textual descriptions, representing the \textcolor{Mahogany}{\textbf{Visual}} and \textcolor{RoyalBlue}{\textbf{Textual}} modalities in our recommendation scenario. At the time of this submission, we have 6,386 valid items, 19,440 users, and 138,821 recorded user-item interactions.

\subsection{Multimodal feature extraction}

Our benchmarking analysis covers three multimodal feature extraction settings. Firstly, in a common multimodal recommendation scenario, we extract \textcolor{Mahogany}{\textbf{Visual}} and \textcolor{RoyalBlue}{\textbf{Textual}} features using pre-trained ResNet50 and SentenceBert, resulting in 2048- and 768-dimensional embeddings, respectively. Secondly, we introduce a novel feature in \framework by employing a pre-trained multimodal model (CLIP) with ViT-B/16, producing 512-dimensional embeddings for both modalities. Thirdly, we showcase another \framework functionality: the extraction of multimodal features through a pre-trained custom model, MMFashion\footnote{\url{https://drive.google.com/open?id=1LmC4aKiOY3qmm9qo6RNDU5v_o-xDCAdT}.}~\cite{DBLP:conf/mm/LiuLWL21}. For the \textcolor{Mahogany}{\textbf{Visual}} modality, it uses a fine-tuned ResNet50 backbone for fashion attribute prediction, yielding 2048-dimensional embeddings. For the \textcolor{RoyalBlue}{\textbf{Textual}} modality, we adopt SentenceBert, resulting in 768-dimensional embeddings.

\subsection{Multimodal recommendation}
We use three multimodal recommendation approaches: VBPR~\cite{DBLP:conf/aaai/HeM16, DBLP:conf/mm/Zhang00WWW21}, BM3~\cite{DBLP:conf/www/ZhouZLZMWYJ23}, and FREEDOM~\cite{DBLP:conf/mm/ZhouS23}. We train and test the three recommendation models by adopting their re-implementations within Elliot~\cite{DBLP:conf/sigir/AnelliBFMMPDN21}; you may consider this public repository\footnote{\url{https://github.com/sisinflab/Formal-MultiMod-Rec}.} for a reference of the models' codes and settings. Technically, we start from the whole user-item interaction data and perform a 80\%/20\% random hold-out split for the training and test sets, retaining the 10\% of the training as validation. Then, we perform a grid search hyper-parameter exploration for each model by following the same experimental setting proposed in~\cite{DBLP:journals/corr/abs-2309-05273}, where the Recall@20 is used as a validation metric. Note that each of the three multimodal recommendation approaches is trained and tested all over again for each multimodal feature extraction setting as reported above.

\subsection{Running the multimodal benchmarks}
We provide three ways to run the multimodal benchmarks: (i) locally, through the GitHub repository; (ii) on Google Colab, through a well-documented Jupyter Notebook\footnote{\url{http://tinyurl.com/mpvv39rp}.}; (iii) by instantating and running a Docker container through \ducho's Docker image\footnote{\url{https://hub.docker.com/repository/docker/sisinflabpoliba/ducho/}.} and a new Docker image equipped with Elliot for multimodal recommendation\footnote{\url{https://hub.docker.com/repository/docker/sisinflabpoliba/mm-recsys/}.}. We especially suggest following either (ii) or (iii) since they require few installation steps. Comprehensive documentation is accessible online\footnote{\url{https://github.com/sisinflab/Ducho/tree/main/demos/demo_recsys/README.md}.}. 

\subsection{Results}
\Cref{tab:benchmarking} shows recommendation results in the three multimodal settings reported above on the retrieved Amazon Baby dataset. Note that all considered metrics account for top-20 recommendation lists.

Overall, we notice a general trend across all multimodal feature settings, namely, more recent recommendation approaches perform better than previous solutions in the literature (FREEDOM and BM3 always outperform VBPR on all the recommendation measures). Then, on a finer-grained evaluation regarding each multimodal feature setting, we observe that the adoption of pre-trained multimodal extractors (i.e., CLIP) or vision deep networks fine-tuned on the fashion domain (i.e., MMFashion) may provide improved recommendation performance in some cases. Specifically, compared to the standard setting including ResNet50 and SentenceBert as extractors, VBPR and BM3 generally reach higher accuracy metrics when leveraging CLIP as multimodal feature extractor (for VBPR, this has been shown in~\cite{DBLP:journals/corr/abs-2310-20343}) or MMFashion for the \textcolor{Mahogany}{\textbf{Visual}} modality.

However, the observed trends regarding multimodal feature extractors may not be easily generalized; for instance, there are settings when we recognize a drop in performance with respect to the standard multimodal setting. For instance, consider the consistent performance drop of FREEDOM when using CLIP. Such results call for more careful analyses regarding the possible impact of recent multimodal large models used as feature extractors in multimodal recommendation~\cite{DBLP:journals/corr/abs-2310-20343}. Indeed, this novel evaluation is out of the scope of this work, but we deem \framework may be efficiently and effectively utilized to conduct extensive benchmarking analyses. 

\begin{table}[!t]
\caption{Recommendation results with varying multimodal feature extractors on top-20 recommendation lists.}\label{tab:benchmarking}
\centering
\begin{adjustbox}{width=\columnwidth,center}
\begin{tabular}{llcccc}
\toprule
\textbf{Extractors} & \textbf{Models}
& \textbf{Recall} & \textbf{Precision} & \textbf{nDCG} & \textbf{HR} \\ \cmidrule{1-6}
\multirow{3}{*}{\makecell[l]{\textcolor{Mahogany}{\textbf{Visual}}\textbf{:} ResNet50\\\textcolor{RoyalBlue}{\textbf{Textual}}\textbf{:} SentenceBert}} & VBPR & 0.0613 & 0.0055 & 0.0309 & 0.1031 \\
& BM3 & 0.0850 & 0.0076 & 0.0426 & 0.1420 \\
& FREEDOM & 0.0935 & 0.0083 & 0.0465 & 0.1537 \\
\cmidrule{1-6}
\multirow{3}{*}{\textcolor{Mahogany}{\textbf{Visual}} \textbf{\&} \textcolor{RoyalBlue}{\textbf{Textual}}\textbf{:} CLIP} & VBPR & 0.0630 & 0.0057 & 0.0313 & 0.1068 \\
& BM3 & 0.0851 & 0.0076 & 0.0428 & 0.1425 \\
& FREEDOM & 0.0704 & 0.0064 & 0.0351 & 0.1187 \\
\cmidrule{1-6}
\multirow{3}{*}{\makecell[l]{\textcolor{Mahogany}{\textbf{Visual}}\textbf{:} MMFashion\\\textcolor{RoyalBlue}{\textbf{Textual}}\textbf{:} SentenceBert}} & VBPR & 0.0619 & 0.0055 & 0.0309 & 0.1027 \\
& BM3 & 0.0847 & 0.0076 & 0.0420 & 0.1423\\
& FREEDOM & 0.0932 & 0.0083 & 0.0465 & 0.1542 \\
\bottomrule
\end{tabular}
\end{adjustbox}
\end{table}
\section{Conclusion and Future Work}

This paper introduces \framework, the new version of our framework \duchoOld for the unified extraction and processing of multimodal features for multimodal recommendation. Building upon its predecessor, \framework includes: (i) framework customization and optimization and (ii) the introduction of novel backend-modality settings and large multimodal models. On the one hand, \framework facilitates the integration of personalized extractor models featuring custom extraction layers, along with the application of pre-processing tasks like custom image normalization. To enhance the data loading and storing, we exploit the widely-used PyTorch custom dataloader. On the other hand, we expand the suite of available backend-modality configurations and implement multimodal-by-design extraction models, allowing modality fusion. To test the new functionalities, we use \framework on top of a multimodal recommendation framework, opening to extensive benchmarking analyses. We plan to further enhance the framework's optimization through multi-GPU extractions and integrate other multimodal-by-design models. Moreover, we intend to use \framework to bring our contribution to multimodal recommendation.

\begin{acks}
This work has been carried out while \textit{Matteo Attimonelli} was enrolled in the Italian National Doctorate on Artificial Intelligence run by Sapienza University of Rome in collaboration with \textit{Politecnico Di Bari}. This work was partially supported by the following projects: CT\_FINCONS\_III, OVS Fashion Retail Reloaded, LUTECH DIGITALE 4.0, VAI2C, IDENTITA, REACH-XY. We acknowledge the CINECA award under the ISCRA initiative, for the availability of high-performance computing resources and support.
\end{acks}

\bibliographystyle{ACM-Reference-Format}
\bibliography{bibliography}


\begin{thebibliography}{10}


\ifx \showCODEN    \undefined \def \showCODEN     #1{\unskip}     \fi
\ifx \showDOI      \undefined \def \showDOI       #1{#1}\fi
\ifx \showISBNx    \undefined \def \showISBNx     #1{\unskip}     \fi
\ifx \showISBNxiii \undefined \def \showISBNxiii  #1{\unskip}     \fi
\ifx \showISSN     \undefined \def \showISSN      #1{\unskip}     \fi
\ifx \showLCCN     \undefined \def \showLCCN      #1{\unskip}     \fi
\ifx \shownote     \undefined \def \shownote      #1{#1}          \fi
\ifx \showarticletitle \undefined \def \showarticletitle #1{#1}   \fi
\ifx \showURL      \undefined \def \showURL       {\relax}        \fi
\providecommand\bibfield[2]{#2}
\providecommand\bibinfo[2]{#2}
\providecommand\natexlab[1]{#1}
\providecommand\showeprint[2][]{arXiv:#2}

\bibitem[Anelli et~al\mbox{.}(2021)]%
        {DBLP:conf/sigir/AnelliBFMMPDN21}
\bibfield{author}{\bibinfo{person}{Vito~Walter Anelli}, \bibinfo{person}{Alejandro Bellog{\'{\i}}n}, \bibinfo{person}{Antonio Ferrara}, \bibinfo{person}{Daniele Malitesta}, \bibinfo{person}{Felice~Antonio Merra}, \bibinfo{person}{Claudio Pomo}, \bibinfo{person}{Francesco~Maria Donini}, {and} \bibinfo{person}{Tommaso~Di Noia}.} \bibinfo{year}{2021}\natexlab{}.
\newblock \showarticletitle{Elliot: {A} Comprehensive and Rigorous Framework for Reproducible Recommender Systems Evaluation}. In \bibinfo{booktitle}{\emph{{SIGIR}}}. \bibinfo{publisher}{{ACM}}, \bibinfo{pages}{2405--2414}.
\newblock


\bibitem[He and McAuley(2016)]%
        {DBLP:conf/aaai/HeM16}
\bibfield{author}{\bibinfo{person}{Ruining He} {and} \bibinfo{person}{Julian~J. McAuley}.} \bibinfo{year}{2016}\natexlab{}.
\newblock \showarticletitle{{VBPR:} Visual Bayesian Personalized Ranking from Implicit Feedback}. In \bibinfo{booktitle}{\emph{{AAAI}}}. \bibinfo{publisher}{{AAAI} Press}, \bibinfo{pages}{144--150}.
\newblock


\bibitem[Liu et~al\mbox{.}(2021)]%
        {DBLP:conf/mm/LiuLWL21}
\bibfield{author}{\bibinfo{person}{Xin Liu}, \bibinfo{person}{Jiancheng Li}, \bibinfo{person}{Jiaqi Wang}, {and} \bibinfo{person}{Ziwei Liu}.} \bibinfo{year}{2021}\natexlab{}.
\newblock \showarticletitle{MMFashion: An Open-Source Toolbox for Visual Fashion Analysis}. In \bibinfo{booktitle}{\emph{{ACM} Multimedia}}. \bibinfo{publisher}{{ACM}}, \bibinfo{pages}{3755--3758}.
\newblock


\bibitem[Malitesta et~al\mbox{.}(2023a)]%
        {DBLP:journals/corr/abs-2309-05273}
\bibfield{author}{\bibinfo{person}{Daniele Malitesta}, \bibinfo{person}{Giandomenico Cornacchia}, \bibinfo{person}{Claudio Pomo}, \bibinfo{person}{Felice~Antonio Merra}, \bibinfo{person}{Tommaso~Di Noia}, {and} \bibinfo{person}{Eugenio~Di Sciascio}.} \bibinfo{year}{2023}\natexlab{a}.
\newblock \showarticletitle{Formalizing Multimedia Recommendation through Multimodal Deep Learning}.
\newblock \bibinfo{journal}{\emph{CoRR}}  \bibinfo{volume}{abs/2309.05273} (\bibinfo{year}{2023}).
\newblock


\bibitem[Malitesta et~al\mbox{.}(2023b)]%
        {DBLP:conf/mm/MalitestaGPN23}
\bibfield{author}{\bibinfo{person}{Daniele Malitesta}, \bibinfo{person}{Giuseppe Gassi}, \bibinfo{person}{Claudio Pomo}, {and} \bibinfo{person}{Tommaso~Di Noia}.} \bibinfo{year}{2023}\natexlab{b}.
\newblock \showarticletitle{Ducho: {A} Unified Framework for the Extraction of Multimodal Features in Recommendation}. In \bibinfo{booktitle}{\emph{{ACM} Multimedia}}. \bibinfo{publisher}{{ACM}}, \bibinfo{pages}{9668--9671}.
\newblock


\bibitem[Radford et~al\mbox{.}(2021)]%
        {DBLP:conf/icml/RadfordKHRGASAM21}
\bibfield{author}{\bibinfo{person}{Alec Radford}, \bibinfo{person}{Jong~Wook Kim}, \bibinfo{person}{Chris Hallacy}, \bibinfo{person}{Aditya Ramesh}, \bibinfo{person}{Gabriel Goh}, \bibinfo{person}{Sandhini Agarwal}, \bibinfo{person}{Girish Sastry}, \bibinfo{person}{Amanda Askell}, \bibinfo{person}{Pamela Mishkin}, \bibinfo{person}{Jack Clark}, \bibinfo{person}{Gretchen Krueger}, {and} \bibinfo{person}{Ilya Sutskever}.} \bibinfo{year}{2021}\natexlab{}.
\newblock \showarticletitle{Learning Transferable Visual Models From Natural Language Supervision}. In \bibinfo{booktitle}{\emph{{ICML}}} \emph{(\bibinfo{series}{Proceedings of Machine Learning Research}, Vol.~\bibinfo{volume}{139})}. \bibinfo{publisher}{{PMLR}}, \bibinfo{pages}{8748--8763}.
\newblock


\bibitem[Yi et~al\mbox{.}(2023)]%
        {DBLP:journals/corr/abs-2310-20343}
\bibfield{author}{\bibinfo{person}{Zixuan Yi}, \bibinfo{person}{Zijun Long}, \bibinfo{person}{Iadh Ounis}, \bibinfo{person}{Craig Macdonald}, {and} \bibinfo{person}{Richard McCreadie}.} \bibinfo{year}{2023}\natexlab{}.
\newblock \showarticletitle{Large Multi-modal Encoders for Recommendation}.
\newblock \bibinfo{journal}{\emph{CoRR}}  \bibinfo{volume}{abs/2310.20343} (\bibinfo{year}{2023}).
\newblock


\bibitem[Zhang et~al\mbox{.}(2021)]%
        {DBLP:conf/mm/Zhang00WWW21}
\bibfield{author}{\bibinfo{person}{Jinghao Zhang}, \bibinfo{person}{Yanqiao Zhu}, \bibinfo{person}{Qiang Liu}, \bibinfo{person}{Shu Wu}, \bibinfo{person}{Shuhui Wang}, {and} \bibinfo{person}{Liang Wang}.} \bibinfo{year}{2021}\natexlab{}.
\newblock \showarticletitle{Mining Latent Structures for Multimedia Recommendation}. In \bibinfo{booktitle}{\emph{{ACM} Multimedia}}. \bibinfo{publisher}{{ACM}}, \bibinfo{pages}{3872--3880}.
\newblock


\bibitem[Zhou and Shen(2023)]%
        {DBLP:conf/mm/ZhouS23}
\bibfield{author}{\bibinfo{person}{Xin Zhou} {and} \bibinfo{person}{Zhiqi Shen}.} \bibinfo{year}{2023}\natexlab{}.
\newblock \showarticletitle{A Tale of Two Graphs: Freezing and Denoising Graph Structures for Multimodal Recommendation}. In \bibinfo{booktitle}{\emph{{ACM} Multimedia}}. \bibinfo{publisher}{{ACM}}, \bibinfo{pages}{935--943}.
\newblock


\bibitem[Zhou et~al\mbox{.}(2023)]%
        {DBLP:conf/www/ZhouZLZMWYJ23}
\bibfield{author}{\bibinfo{person}{Xin Zhou}, \bibinfo{person}{Hongyu Zhou}, \bibinfo{person}{Yong Liu}, \bibinfo{person}{Zhiwei Zeng}, \bibinfo{person}{Chunyan Miao}, \bibinfo{person}{Pengwei Wang}, \bibinfo{person}{Yuan You}, {and} \bibinfo{person}{Feijun Jiang}.} \bibinfo{year}{2023}\natexlab{}.
\newblock \showarticletitle{Bootstrap Latent Representations for Multi-modal Recommendation}. In \bibinfo{booktitle}{\emph{{WWW}}}. \bibinfo{publisher}{{ACM}}, \bibinfo{pages}{845--854}.
\newblock


\end{thebibliography}

\end{document}